\documentclass[twocolumn,english,aps,prl,showpacs]{revtex4}
\usepackage{amsmath,graphicx,amssymb,epsfig,babel,dsfont}

\renewcommand{\Im}{{\rm Im}}

\newcommand{\rD}{{\rm D}}
\newcommand{\rS}{{\rm S}}
\newcommand{\rG}{{\rm G}}
\newcommand{\rE}{{\rm E}}
\newcommand{\rF}{{\rm F}}
\newcommand{\rs}{{\rm s}}
\newcommand{\rp}{{\rm p}}
\newcommand{\rc}{{\rm c}}
\newcommand{\kB}{k_{\rm B}}

\begin{document}

\title{Towards boolean operations with thermal photons}

\author{Philippe Ben-Abdallah}
\email{pba@institutoptique.fr}
\affiliation{Laboratoire Charles Fabry,UMR 8501, Institut d'Optique, CNRS, Universit\'{e} Paris-Sud 11,
2, Avenue Augustin Fresnel, 91127 Palaiseau Cedex, France}
\affiliation{Universit\'{e} de Sherbrooke, Department of Mechanical Engineering, Sherbrooke, PQ J1K 2R1, Canada.}

\author{Svend-Age Biehs}
\email{s.age.biehs@uni-oldenburg.de}
\affiliation{Institut f\"{u}r Physik, Carl von Ossietzky Universit\"{a}t, D-26111 Oldenburg, Germany.}

\date{\today}


\begin{abstract}
The Boolean algebra is the natural theoretical framework for a classical information treatment. The basic logical operations are 
usually performed using logic gates. In this Letter we demonstrate that NOT, OR and AND gates can be realized exploiting the near-field
radiative interaction in N-body systems with phase change materials. With the recent development of a photon thermal transistor and thermal memory,  this result 
paves the way for a full information treatment and smart solutions for active thermal management at nanoscale with photons.  
\end{abstract}

\maketitle

Controlling heat exchanges at nanoscale is a tremendeous challenge for the developement of numerous future technologies. During the 
last decade, thermal analogs of diodes, memories and transistors have been introduced~\cite{Casati1,BaowenLiEtAl2012} in order to 
control (switch, modulate, store and even amplify), the flow of heat and energy carried by phonons in solid elements networks in the same
manner as the flow of electrons is controlled in electric circuits. In 2007, thermal logic gates~\cite{BaowenLi2} have been proposed 
to build up thermal logic calculations using the transport of acoustic phonons in lattices of nonlinear solid elements. 
Recently, to overcome the problems linked to the relatively small speed of acoustic phonons and due to the presence of localized 
Kapitza resistances inside these lattices, radiative analogs of diodes~\cite{OteyEtAl2010,Basu,Fan,PBA_APL,NefzaouiEtAl2013},
transistor \cite{PBA_PRL2014,Joulain2015}, memory \cite{Kubytskyi},  splitters~\cite{PBAEtAl2015,PBA2016} and 
even electric wire~\cite{MessinaEtAl2016} have been proposed for controling radiative heat currents both in near and far-field regimes in 
complex architectures of solids out of contact opening so the way to contactless thermal analogs of basic electronic devices. Some of
these theoretical concepts  --- the diode and the memory --- have been verified experimentally very recently~\cite{ItoEtAl2015,ItoEtAl2016}.

In this Letter we introduce the concept of photonic thermal logic gates which use thermal photons instead of electrons to make logical operations. 
This element is the basic building block for future thermal circuits which implement Boolean functions by performing logical operations on single or several logical 
inputs in order to produce a single logical output. In a thermal logic gate, a power is supplied to some points of the device which function as inputs 
to keep them at specific temperatures and to bring another point of the device which functions as the output to a certain temperature level. By 
defining temperature levels which serve as tresholds to define logical '0' or '1', a truth table can be associated to the inputs and the output to define specific 
logical operations. 

\begin{figure}[Hhbt]
\includegraphics[scale=0.35]{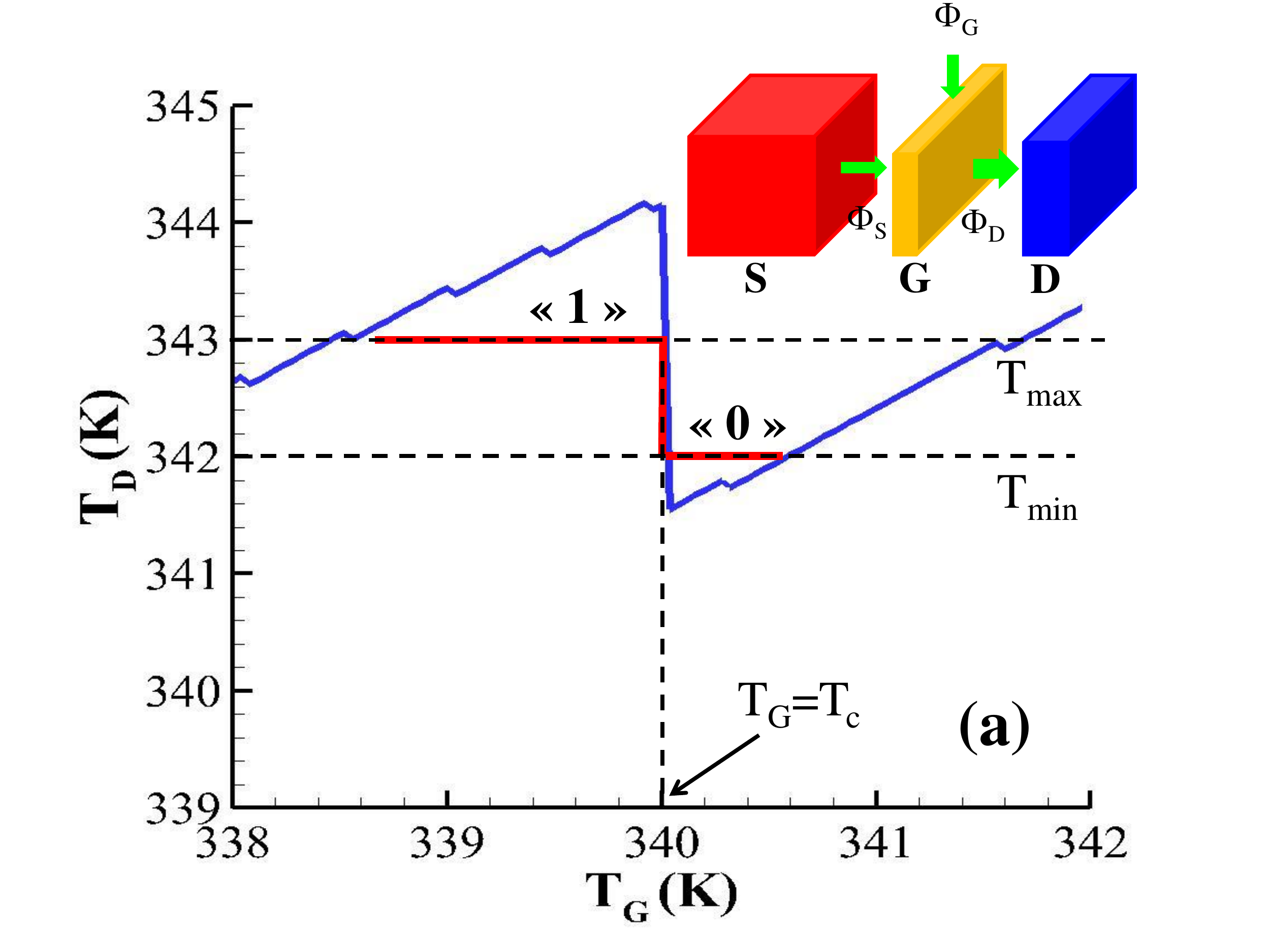}
\includegraphics[scale=0.25]{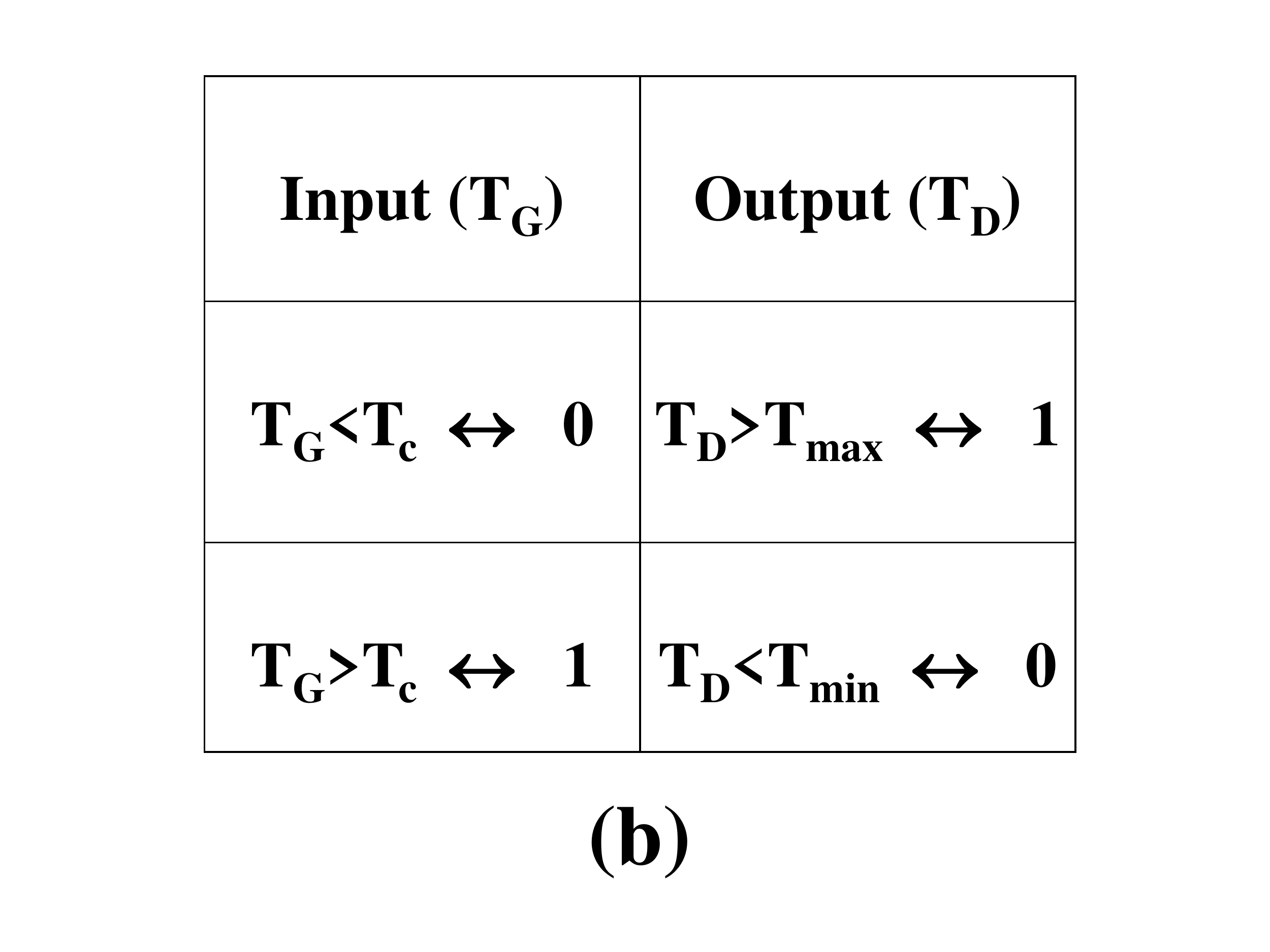}
\caption{(a) Radiative NOT gate made with a SiO$_2$/VO$_2$/SiO$_2$ plane thermal transistor. The source (S) is assumed semi-infinite while the gate (G) and the drain (D) have a tkickness of 100 nm. The separation  distance are d=100 nm. The temperature $T_G$ of the gate define the input and the drain temperature sets the gate output. In the operating range of the gate around the critical temperature $T_c$ of the gate, if $T_D<T_{min}$ the gate is in the thermal state "0". On the contrary,  when $T_D>T_{max}$ the gate is in the state "1". The rectangular function represents the ideal response of the gate in its operating range.(b) Thruth table for the ideal NOT gate.
\label{NOT gate}}
\end{figure}

To start, let us consider the simplest logic gate, the NOT gate which implements a logical negation. This operation can be performed by using 
a three body system as sketched in Fig.~1-a. It has been shown that this system consisting of a 
thermal reservoir acting as the heat source, an intermediate gate layer made of an insulator-metal transition (IMT) material~\cite{Mott,Baker} and a second reservoir which serves as a heat drain is a thermal transistor~\cite{PBA_PRL2014}. It has been demonstrated by us~\cite{PBA_PRL2014} that due to the ability of the gate layer to qualitatively and quantitatively change its optical properties through a small change of its temperature around a critical temperature $T_\rc$ the heat flux towards the drain can be switched, modulated and even amplified. In order to operate the thermal transistor as a NOT gate the
temperature of the source is maintained at a fixed temperature $T_{\rm S}$. Throughout the manuscript we use $T_{\rm S} = 360\,{\rm K}$. 
Then, the intermediate layer functions as the input of the NOT gate and the drain as the output. That means the temperature $T_{\rm G}$ of the gate layer which 
is assumed to be smaller than  $T_{\rm S}$ sets the boolean input of the NOT gate. Here we define the thermal state with $T_{\rm G} < T_{\rm c}$ as '0' and the 
termal state with  $T_{\rm G} > T_{\rm c}$ as '1'. This state can be set from outside by adding or removing heat from the intermediate gate layer. 
Finally, since $T_\rS$ and $T_\rG$ are fixed by an external reservoir there will be a radiative energy flux $\Phi_\rD$ from the source and the
gate towards the drain until $T_\rD$ has reached a value such that 
\begin{equation}
  \Phi_\rD(T_\rS,T_\rG,T_\rD) = 0,
\label{Temp}
\end{equation}
i.e.\ the drain is in its local thermal equilibrium and the system has reached the steady state. Now, we can readout the output temperature $T_\rD$ which defines 
the boolean output of the NOT gate. Here we define the thermal state '0' as the state where $T_\rD < T_{\rm min}$ and the termal state '1' 
as the state where $T_\rD > T_{\rm max}$ introducing two thresholds $T_{\rm min}$ and $T_{\rm max}$ with $T_{\rm min}<T_{\rm max}$. Here
we choose $T_{\rm min} = 342 K$ and $T_{\rm max} = 343 K$.

In order to verify if the transistor operates as a NOT gate we need to determine $T_\rD$ for different 
input values of $T_\rG$ by evaluating the roots of Eq.~(\ref{Temp}). For a system operating in near-field regime (i.e.\ far-field exchanges with the surrounding can be neglected) the flux received by photon tunneling by the drain reads~\cite{Messina}
\begin{equation}
  \Phi_{\rD}= \int_0^\infty\!\frac{d\omega}{2\pi}\,\phi_\rD(\omega,d), 
\label{Eq:Flux_D}
\end{equation}
where the monochromatic heat flux is given by
\begin{equation}
\begin{split}
  \phi_{\rD} &= \hbar\omega\sum_{j = \{\rm s,p\}}\int\! \frac{{\rm d}^2 \boldsymbol{\kappa}}{(2 \pi)^2} \, \bigl[n_{\rS\rG}(\omega)\mathcal{T}^{\rS/\rG}_j(\omega,\boldsymbol{\kappa}; d)\\
                &\quad+n_{\rG\rD}(\omega)\mathcal{T}^{\rG/\rD}_j(\omega,\boldsymbol{\kappa}; d)\bigr].
\end{split}
\label{Eq:Flux_D_1}
\end{equation}

\begin{figure}[Hhbt]
\includegraphics[scale=0.38]{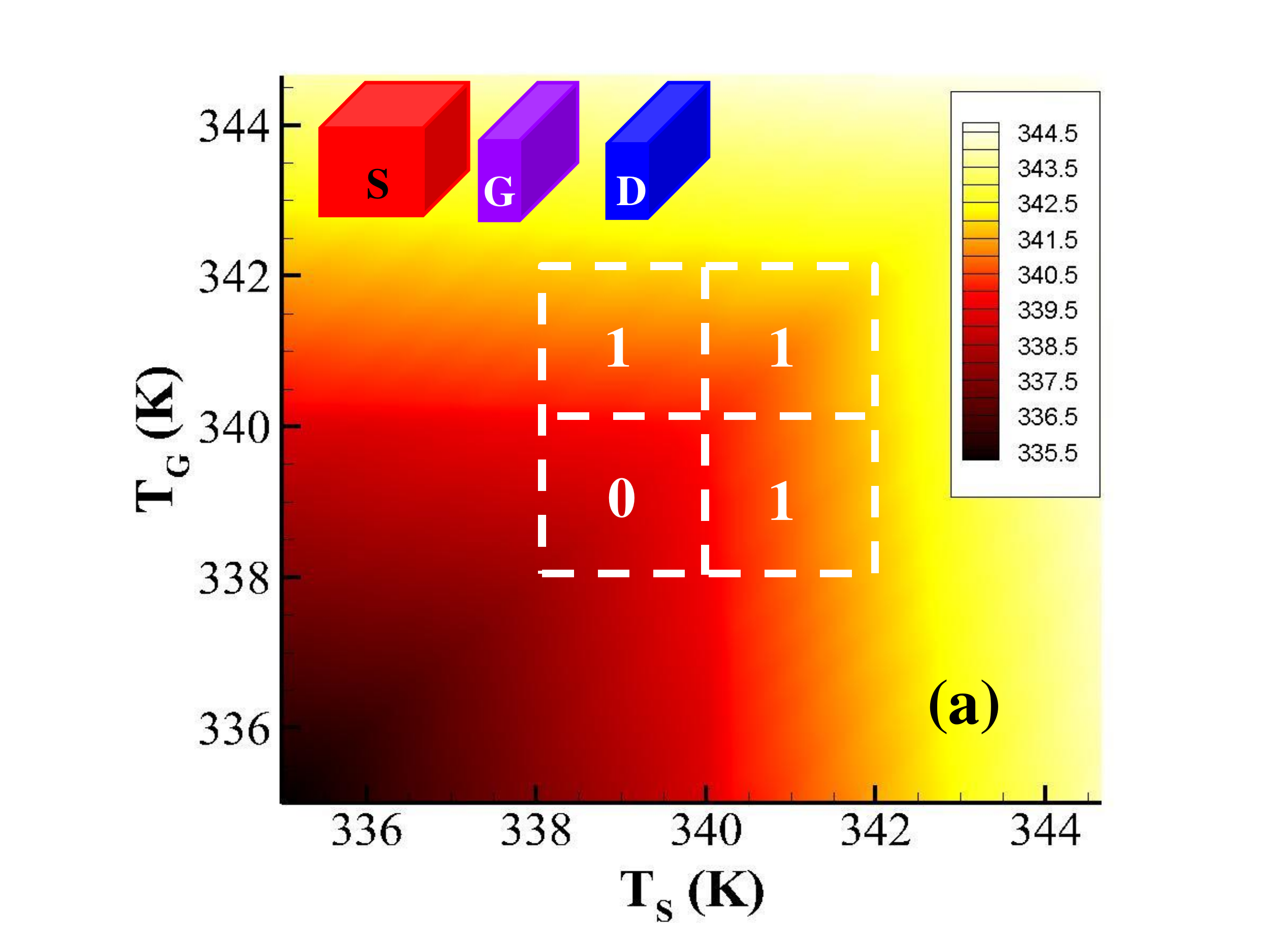}
\includegraphics[scale=0.3]{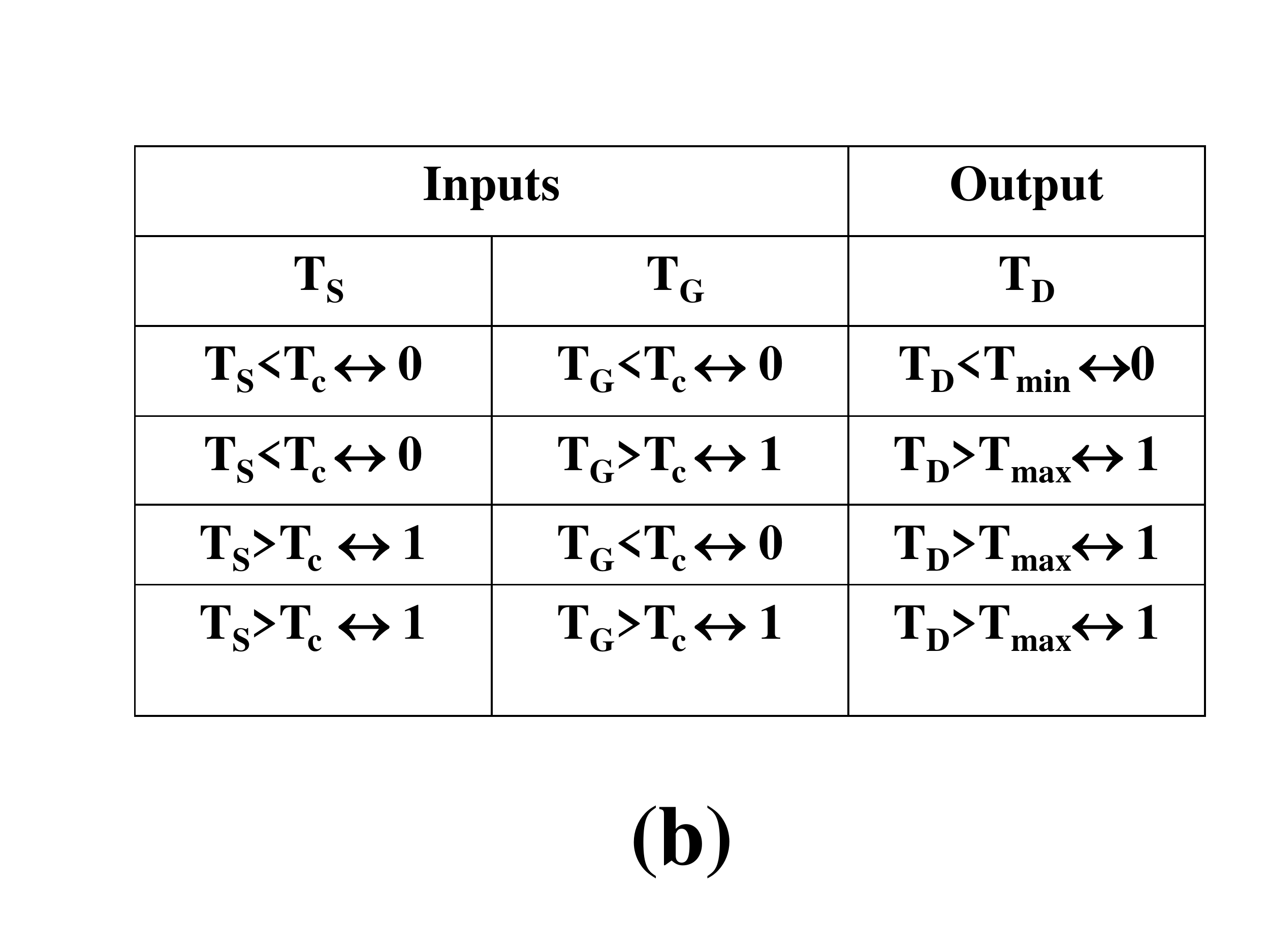}
\caption{(a) Radiative OR gate made with a  VO$_2$/VO$_2$ /SiO$_2$  thermal transistor. The source (S) is assumed semi-infinite while the  gate (G) and the drain (D) have a tkickness of 100 nm. The separation  distances are d=100 nm. The two temperatures $T_{S}$ and  $T_{G}$ define the inputs and the drain temperature sets the gate output. The operating range of the OR gate is delimited by dashed rectangular domain centered at $(T_{S},T_{G})=(T_c,T_c)$  with $T_c=340 K$ .  (b) Thruth table for the ideal OR gate.
\label{OR gate}}
\end{figure}

Here $\mathcal{T}^{\rS/\rG}_j \in [0,1]$ and $\mathcal{T}^{\rG/\rD}_j \in [0,1]$ denote the efficiencies of coupling of each mode $(\omega,\boldsymbol{\kappa})$ between the source and the gate 
and between the gate and the drain for both polarization states $j = \rs, \rp$; $\boldsymbol{\kappa} = (k_x,k_y)^t$ is the wavevector parallel to the surfaces of the multilayer
system. In the above relation $n_{ij}$ denotes the difference of Bose-distribution functions $n_i$ and $n_j$ [with $n_{i/j}= (\exp\bigl({\frac{\hbar \omega}{\kB T_{i/j}}}\bigr)-1)^{-1}$] 
at the frequency $\omega$; $\kB$ is Boltzmann's constant and $2 \pi \hbar$ is Planck's constant. According to the N-body near-field heat transfer theory presented in Ref.~\cite{Messina}, 
the transmission coefficients $\mathcal{T}^{\rS/\rG}_j$ and $\mathcal{T}^{\rG/\rD}_j$ of the energy carried by each mode written in terms of optical reflection 
coefficients $\rho_{\rE,j}$ ($\rE = \rS, \rD, \rG$) and transmission coefficients $\tau_{\rE,j}$ of each basic element of the system and in terms of reflection 
coefficients $\rho_{\rE\rF,j}$ of couples of elementary elements~\cite{Messina}
\begin{equation}
\begin{split}
  &\mathcal{T}^{\rS/\rG}_{j}(\omega,\boldsymbol{\kappa},d)\\
  &=\frac{4\mid\tau_{\rG,j}\mid^2 \Im(\rho_{\rS,j})\Im(\rho_{\rD,j})e^{-4\gamma d}}{\mid 1-\rho_{\rS\rG,j}\rho_{\rD,j}e^{-2\gamma d}\mid^2\mid1-\rho_{\rS,j}\rho_{\rG,j}e^{-2\gamma d}\mid^2},\\
  &\mathcal{T}^{\rG/\rD}_{j}(\omega,\boldsymbol{\kappa},d)=\frac{4 \Im(\rho_{\rS\rG,j})\Im(\rho_{\rD,j})e^{-2\gamma d}}{\mid 1-\rho_{\rS\rG,j}\rho_{\rD,j}e^{-2\gamma d}\mid^2}
\end{split}
\label{Trans}
\end{equation}
introducing the imaginary part of  wavevector normal to the surfaces in the multilayer structure $\gamma = \Im(k_z) = \sqrt{\kappa^2 - \omega^2/c^2}$; 
$c$ is the velocity of light in vacuum. For the detailed expressions of the reflection coefficients we refer to Ref.~\cite{Messina}.

By solving the transcendental Eq.~(\ref{Temp}) using expression ~(\ref{Eq:Flux_D}) we obtain the results shown in Fig.~1-a. The configuration of
the NOT gate used in the numerical calculation is as follows: both source and drain are made of silica~\cite{Palik} where the source is assumed to be semi-infinite and
the drain is assumed to have a thickness of $100\,{\rm nm}$. The intermediate $100\,{\rm nm}$ thick gate layer is made of vanadium dioxide (VO$_2$) an IMT material which 
undergoes a first-order transition (Mott transition~\cite{Mott}) from a high-temperature metallic phase to a low-temperature insulating phase~\cite{Baker} at a critical 
temperature $T_{\rm c}$ which is close to room-temperature. Here we assume for convenience that the phase transition happens abruptly at $T_{\rm c}=340\,{\rm K}$ which means
that we are neglecting the phase transition region between the two phases. As demonstrated in previous works~\cite{van Zwol2,van Zwol3,Gotsmann} below its critical 
temperature, i.e.\ for $T_\rG < T_\rc$, VO$_2$  supports surfaces phonon-polariton (SPhP) in the same frequency range as silica. On the contrary, in its metallic phase 
VO$_2$ does not support surface wave resonances anymore so that the near-field interaction between each element in the system is strongly reduced for $T_\rG > T_\rc$. 
Hence, the coupled silica-VO$_2$ system behaves like a thermal switch~\cite{PBA_APL,PBA_PRL2014}. This 'switching' ability can be seen in Fig.~1-a when the temperature
$T_\rD$ makes a sudden jump at $T_\rG = T_\rc$. Finally, we can see in Fig.~1-a that for $T_\rG < T_\rc$ we have a region where $T_\rD > T_{\rm max}$ which means that 
the input state '0' results in the output state '1'. Similarly, for  $T_\rG > T_\rc$ we have a region where  $T_\rD < T_{\rm min}$  which means that the input state '1' 
results in the output state '0'. Hence, the here introduced device functions indeed as a NOT gate with the truth table shown in Fig.~1-b. The performance of the here
introduced NOT gate is close to the idealized NOT gate performance sketched by the red line in Fig.~1-a.

As a second example we sketch the realization of an OR gate using once again a simple gate thermal transistor. However, contrary to the NOT gate we choose here phase change materials for both the source and the gate layer and silica for the drain layer. The temperatures of these two elementary parts of the transistor are usedas inputs while the drain temperature set the output of the logic gate. The temperature evolution of the drain with respect to the temperatures of the source and the gate is plotted in Fig. 2-a.  We see that around the point  $(T_{\rm S},T_{\rm G})=(T_\rc,T_\rc)$  with $T_\rc=340 K$ where the phase change occurs both in the source and the gate, the temperature of the drain undergoes a significant variation.   If $T_{\rm S}$ and $T_{\rm G}$ are both smaller than the critical temperature $T_\rc$ then these two elementary blocks behave like a dielectric so that the temperature of the drain is small. On the contrary, if either the source or the gate undergo a phase change, then the temperature of the drain increases abruptly. Hence, by conveniently intoducing two suitable treshold temperatures $T_{\rm min}$ and $T_{\rm max}$ it is clear that we can associate to the drain two different thermal states with respect to the temperatures $T_\rS$ and $T_\rG$  around the region $(T_{\rm S},T_{\rm G})=(T_\rc,T_\rc)$. Therefore, this transistor behaves like an OR gate with the truth table given in Fig. 2-b. By reversing the definition of thermal states in the drain, it is also clear that this system mimicks a NOR gate.

\begin{figure}[Hhbt]
\includegraphics[scale=0.38]{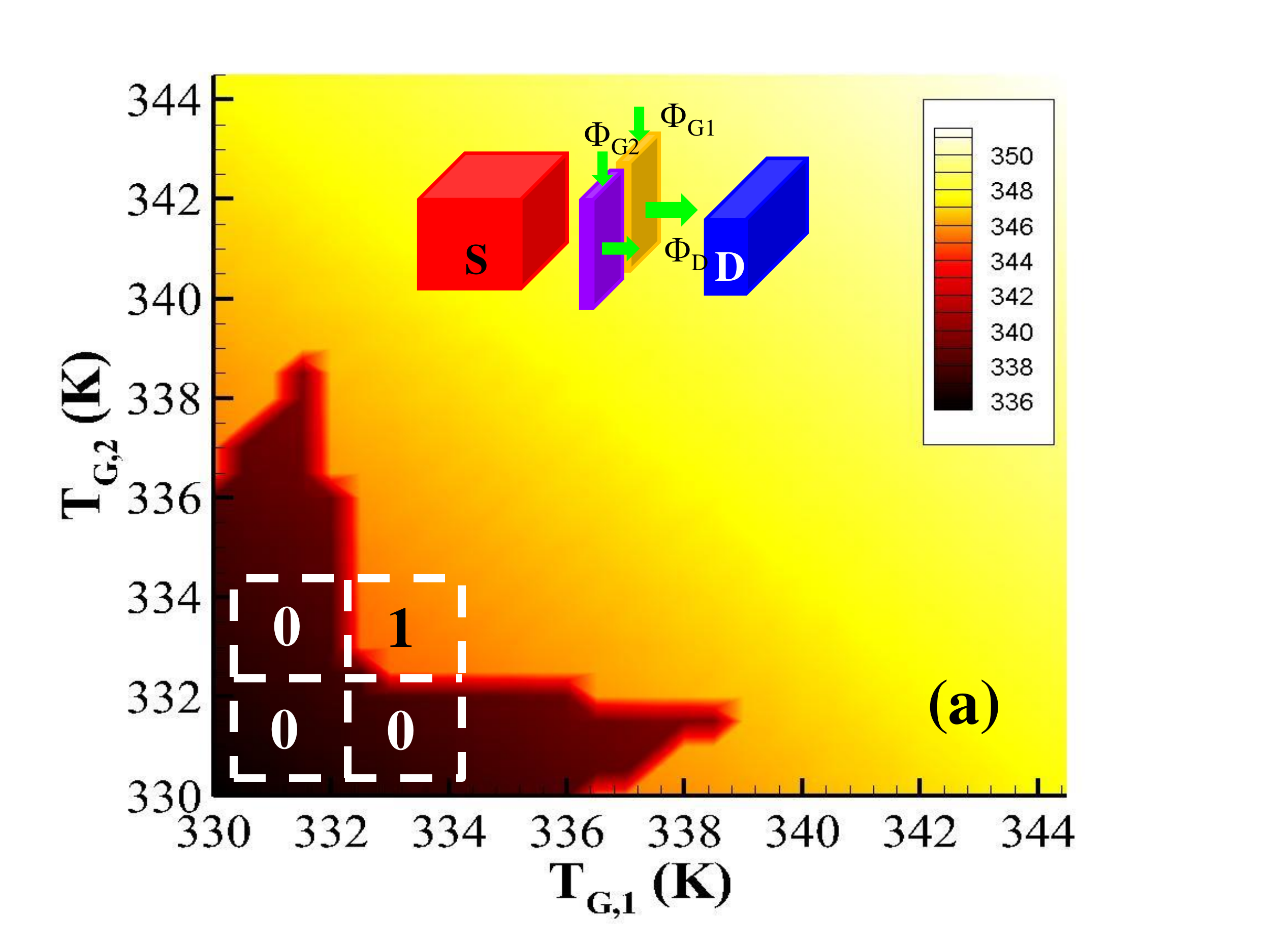}
\includegraphics[scale=0.3]{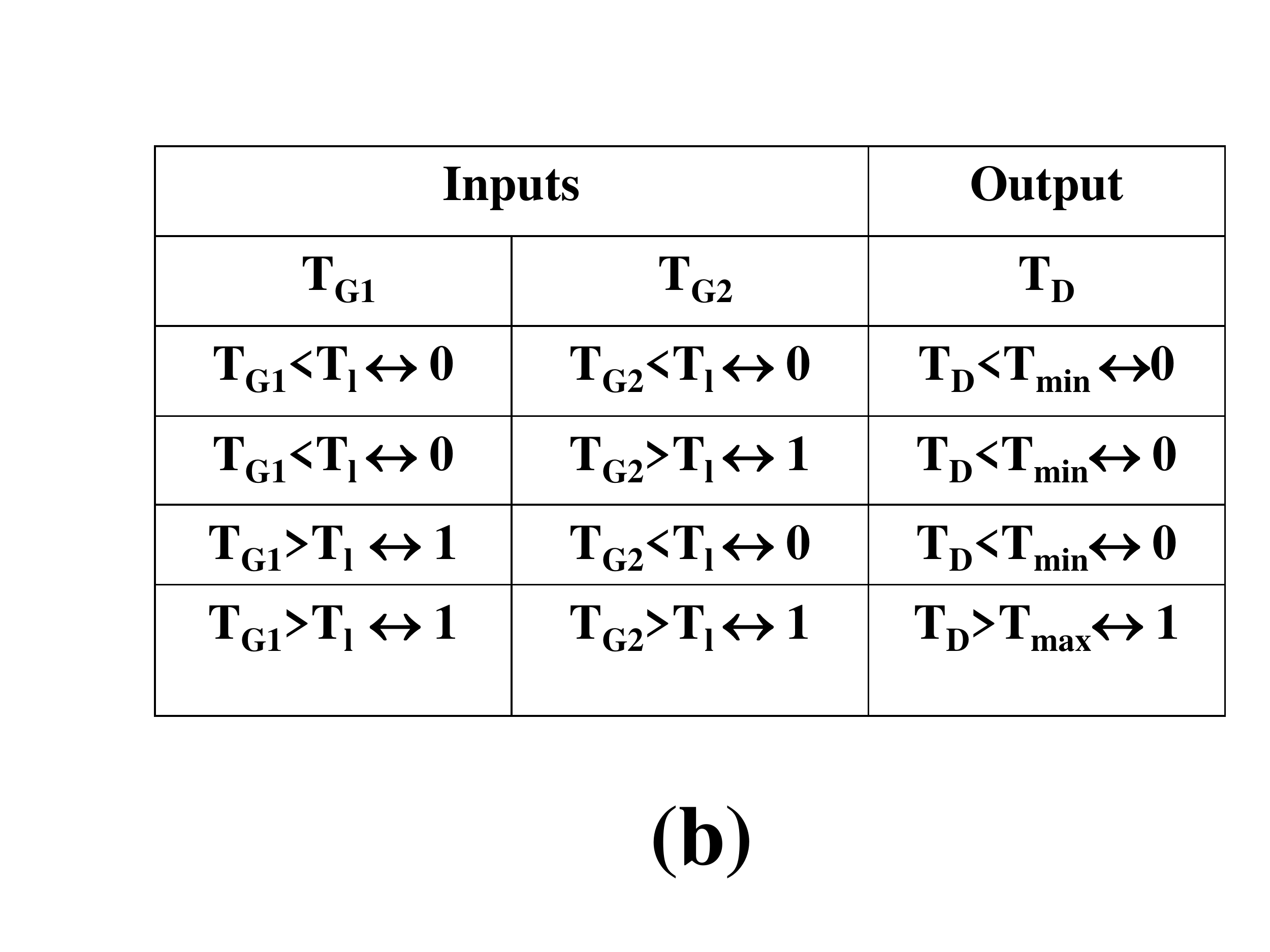}
\caption{(a) Radiative AND gate made with a double SiO$_2$ gate thermal transistor, the source being made in  SiO$_2$ and the drain in VO$_2$. The source (S) is assumed semi-infinite while both gates (G1, G2)  and the drain (D) have a tkickness of 250 nm and 500 nm respectively. The separation  distances are d=100 nm and the gates are assumed isolated one from the other. The two temperatures $T_{G1}$ and  $T_{G2}$ define the inputs and the drain temperature sets the gate output. The operating range of the AND gate is delimited by dashed rectangular domain centered at $(T_{G1},T_{G2})=(T_l,T_l)$  with $T_l=332.2 K$ . (b) Thruth table for the ideal AND gate.
\label{AND gate}}
\end{figure}

Now, we sketch the realization of an AND gate. This double input device is shown in Fig.~3-a. It is a double gate thermal transistor made with two silica gates and a silica source. Contrary to the NOT gate it is the drain which is made of a phase change material. The temperature of both gates set the two inputs of the logic gate. We  assume that the temperature of two gates can now be controlled independent making the assumption, for convenience, that they are thermally insulated one from the other so that we can express the heat flux received by the drain 
just as the mean value of two NOT gates, i.e.\ we have
\begin{equation}
\begin{split}
  \phi_{\rD} &= \frac{\hbar\omega}{2}\sum_{j = \{\rm s,p\}}\int\! \frac{{\rm d}^2 \boldsymbol{\kappa}}{(2 \pi)^2} \, \bigl[n_{\rS\rG1}(\omega)\mathcal{T}^{\rS/\rG1}_j(\omega,\boldsymbol{\kappa}; d)\\
                &\quad+n_{\rG1\rD}(\omega)\mathcal{T}^{\rG1/\rD}_j(\omega,\boldsymbol{\kappa}; d)+n_{\rS\rG2}(\omega)\mathcal{T}^{\rS/\rG2}_j(\omega,\boldsymbol{\kappa}; d)\\
&\quad+n_{\rG2\rD}(\omega)\mathcal{T}^{\rG2/\rD}_j(\omega,\boldsymbol{\kappa}; d)\bigr].
\end{split}
\label{Eq:Flux_D_2}
\end{equation}
Here the transmission coefficients are given by the same expressions as in a single gate transistor. In Fig. 3-a we show the equilibrium temperature $T_\rD$ of drain with respect . In the central region around  $(T_{G1},T_{G2})=(T_l,T_l)$ with $T_l=332.2 K$ we see that $T_\rD$ can undergo a sudden variation after a short change in the gate temperatures. By introducing two critical temperatures $T_{min}$ and  $T_{max}$ we can associate to the drain two thermal states '0' or '1'  with respect the relative value of the drain temperature with respect to these thresholds. As it clearly appears in Fig.3-a this double gate system behaves as a digital AND gate with the thruth table given in Fig. 3-b. Note that by reversing the definition of thermal states '0' and '1' the AND gate becomes a NAND gate.

\begin{figure}[Hhbt]
\includegraphics[scale=0.35]{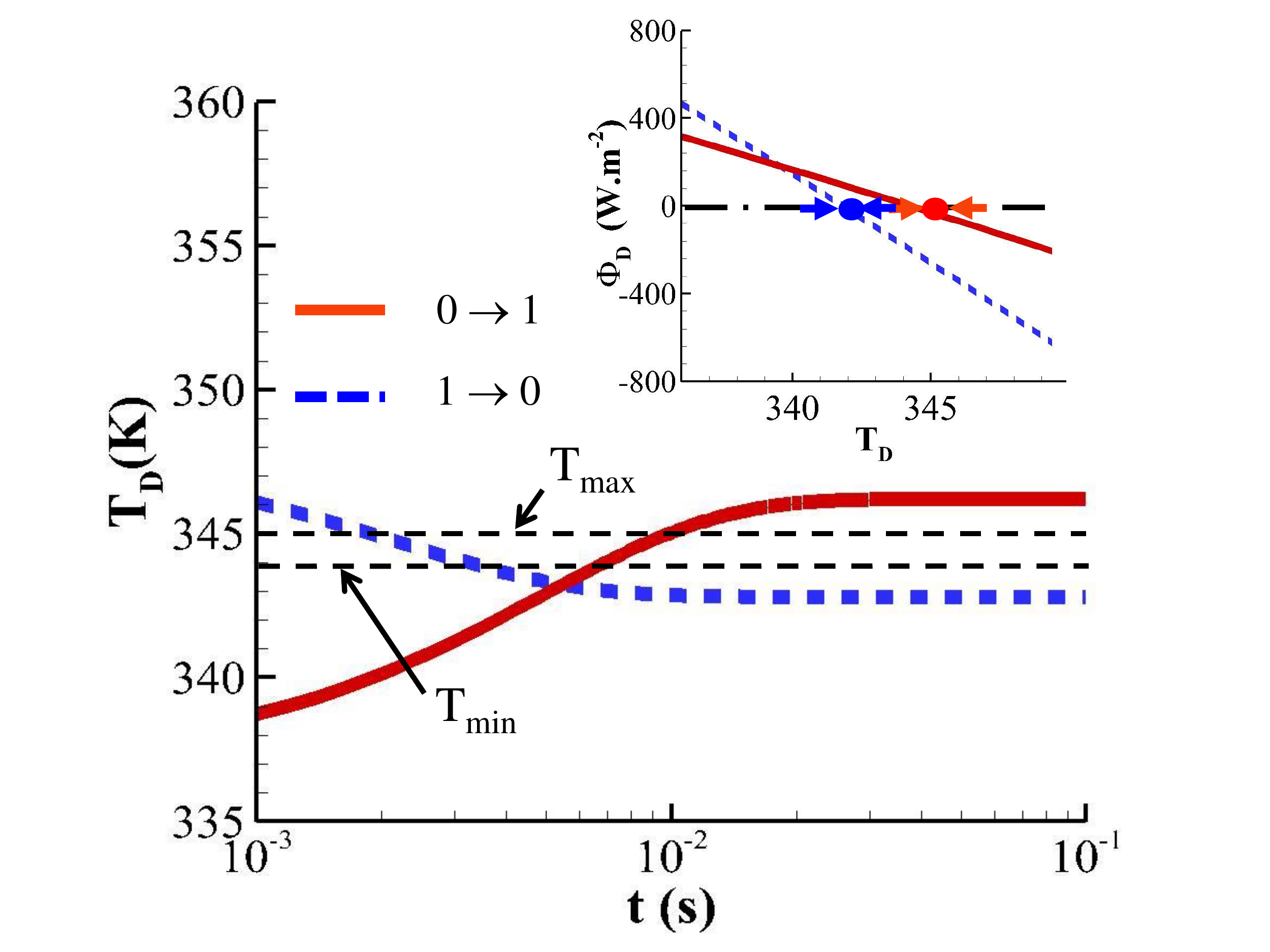}
\caption{Thermal relaxation of the drain  in a SiO$_2$/VO$_2$/SiO$_2$ NOT gate with the geometrical parameters identical as in Fig. 1.  The red solid (resp. blue dashed)  line shows the time evolution of the drain temperature from the state "0" (resp "0") to the state  "1" (resp. "0")  when the initial temperature is $T_D(0)=337 K$ (resp.$T_D(0)=348 K$ ). The gate temperature is assumed fixed at $T_G=343 K$ (resp.  $T_G=337 K$ ) and the consign temperatures are $T_{min}=344 K$ and  $T_{max}=345 K$. The inset shows the stability diagram of solutions for both transitions between the state "0" and the sate "1". 
\label{dyn NOT}}
\end{figure}

To finish we evaluate the time required for these gates to switch from one state to another. To this end we 
study the relaxation process for the drain (100 nm thick) in the SiO$_2$/VO$_2$/SiO$_2$ NOT gate described in Fig.~1. 
The time evolution of the drain temperature from an initial state (set with two different values for $T_D(t=0s)$ corresponding to '0' and '1')  
is determined by the nonlinear dynamic equation
\begin{equation}
\rho C d_\rD \dot{T}_D=\Phi_\rD(T_S,T_G,T_D),
\label{dynamic NOT gate}
\end{equation}
where $\rho$, $C$ and $d_\rD$  denote the mass density, the heat capacity and the thickness of the drain, respectively. In Fig. 4 we show the 
transition dynamic from the state '0' to the state '1'  (from the state '1' to the state '0') with an initial temperature $T_\rD(0)=337 K$ 
($T_\rD(0)=348 K$ ) and a gate at fixed temperature $T_\rG=343 K$ ($T_\rG=337 K$). The overall time the NOT gate takes to switch from state '0' to state '1' (state '1' to state '0') is of few ms. Note that this switching time is relatively large compared to the operating speed of electric logic gates because of the thermal inertia. For even thicker thicker drains this switching time will increase.

To summarize, we have introduced the concept of logic gates for heat radiation by discussing in some details a realization of a NOT, OR and AND gates operating in near-field regime.
The same concepts could obviously be applied in far-field regime that is for separation distances much larger than the thermal wavelength. However, in this regime the magnitude of heat flux is much lower so that the thermalization process is much slower.
The main advantage of radiative thermal logic gates is probably not the possibility of some kind of numerical calculation using heat currents instead of electrical currents, because 
the operation speed is relatively low even in near-field regime, but rather the possibility to actively control temperature distributions and heat flux in dense complex networks made of solid elements which are out of 
contact. The implementation of more complex boolean operations will require, according to the famous De Morgan's law, to combine various logical gates. However, in 
near-field regime this combination cannot be sequential, generally, because of many-body effects which make the heat transport throughout the structure non additive. 
This demands for the development of a general many-body theory in arbitrary solid networks.

%
%

\begin{acknowledgments}
\end{acknowledgments}

\end{document}